\documentstyle[12pt,epsfig]{article}

\newlength{\dinwidth}
\newlength{\dinmargin}
\def\lapproxeq{\lower .7ex\hbox{$\;\stackrel{\textstyle
<}{\sim}\;$}}
\def\gapproxeq{\lower .7ex\hbox{$\;\stackrel{\textstyle
>}{\sim}\;$}}

\def\be{\begin{equation}}
\def\ee{\end{equation}}
\def\bea{\begin{eqnarray}}
\def\eea{\end{eqnarray}}

\def\pp{p\bar{p}}

\def\GeV{{\rm GeV}}

\begin{document}
%\titlepage
\begin{flushright}
IPPP/03/06 \\
DCPT/03/12 \\
12 February 2003 \\
\end{flushright}

\vspace*{2cm}

\begin{center}
{\Large \bf Factorization breaking in diffractive dijet production
%\rule{0mm}{2.5ex}
}

\vspace*{1cm}
\textsc{A.B.~Kaidalov$^{a,b}$, V.A.~Khoze$^{a,c}$, A.D. Martin$^a$ and M.G. Ryskin$^{a,c}$} \\

\vspace*{0.5cm} $^a$ Department of Physics and Institute for
Particle Physics Phenomenology, \\
University of Durham, DH1 3LE, UK \\
$^b$ Institute of Theoretical and Experimental Physics, Moscow, 117259, Russia\\
$^c$ Petersburg Nuclear Physics Institute, Gatchina,
St.~Petersburg, 188300, Russia \\

\end{center}

\vspace*{1cm}

\begin{abstract}
We study diffractive hard dijet production, with one or two
rapidity gaps, at high energies. We emphasize that both hard and
Regge factorization are broken in these processes. We show that a
multi-Pomeron-exchange model for screening effects gives a
specific pattern for the breakdown of factorization, which is in
good agreement with diffractive dijet data collected at the
Tevatron.
\end{abstract}

\vspace*{1cm}

%\section{Introduction}

The investigation of diffractive processes at high energies gives
important information on the structure of hadrons and their
interaction mechanisms. Hard diffractive processes, such as the
diffractive production of dijets, allow the study of the interplay
of small- and large-distance dynamics within QCD. The existence of
a hard scale provides the normalization of the Born term diagram,
which is shown for single-diffractive dissociation in Fig.~1(a)
and for the double-diffractive process\footnote{Here by
double-diffractive we mean a process with two rapidity gaps which,
at high energies, corresponds to double-Pomeron-exchange.} in
Fig.~1(b).
\begin{figure}[!ht]
\begin{center}
\epsfig{figure=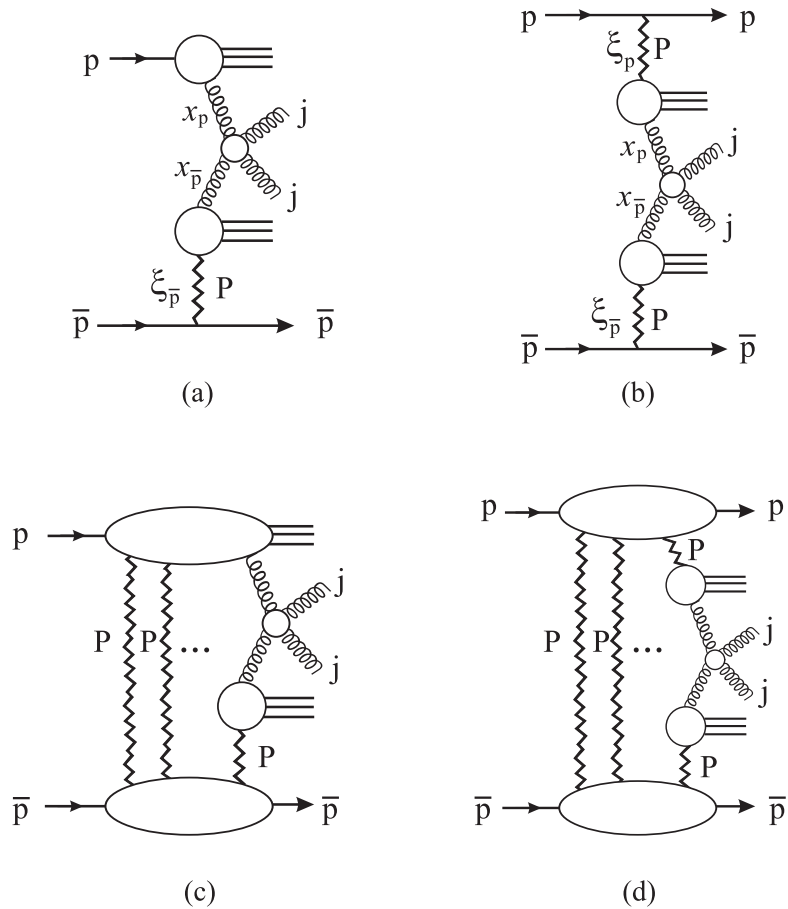,height=5in}\caption{(a,b) The Born
diagrams for diffractive dijet production in high energy $\pp$
collisions with one, two rapidity gaps indicated by Pomeron, P,
exchange.\ (c,d) The multi-Pomeron exchange contributions to the
above processes, where the upper and lower blobs encapsulate all
possible Pomeron permutations.}
\end{center}\end{figure}
These processes are respectively characterized by the existence of
one and two large rapidity gaps, each of which is represented by
Pomeron exchange. At high energies, there are important
contributions from unitarization effects. In the $t$ channel
Reggeon framework, these effects are described by multi-Pomeron
exchange diagrams, Figs~1(c),(d). Such diagrams lead to a strong
violation of both Regge and hard factorization, which were valid
for the Born diagrams of Figs.~1(a),(b).

As has been known for a long time (see, for example, \cite{WK}),
factorization does not necessarily hold for diffractive production
processes; for recent studies see Refs.~\cite{KKMR,B,C} and
references therein. The suppression of the single-Pomeron Born
cross section due to the multi-Pomeron
contributions\footnote{Another approach to the phenomenon of
factorization breaking has been proposed by Goulianos within the
so-called gap probability renormalization model, see~\cite{GOUL}
and references therein.} depends, in general, on the particular
hard process. At the Tevatron energy, $\sqrt s = 1.8$~TeV, the
suppression is in the range 0.05--0.2~\cite{GLM,KMRsoft,BH,KKMR}.
In fact, a computation of this effect was found to give a
quantitative understanding~\cite{KKMR} of the
experimentally-observed suppression of the single diffractive
dijet cross sections at the Tevatron~\cite{CDFfac} as compared to
the predictions based on HERA results~\cite{A9}. The comparison
relies on the partonic distributions in the Pomeron determined
from HERA data. These parton densities have some uncertainty
(especially for the gluonic content of the Pomeron).
Interestingly, when the new fit to the H1 diffractive
data~\cite{SCHILL} is used in the approach of Ref.~\cite{KKMR},
even better agreement with the CDF Tevatron data~\cite{CDFfac} is
achieved~\cite{AMS}.

Nowadays diffractive processes are attracting more attention as a
way of extending the physics programme at proton colliders,
including novel ways of searching for New Physics; see, for
example, \cite{AR,INC,DKMOR}. Clearly, the correct treatment of
the screening effects is crucial for the reliability of the
theoretical predictions of the cross sections for these
diffractive processes. As mentioned above, some tests of the
mechanism of diffractive dijet production have been made
\cite{Liverpool,KKMR,AMS}, but further, model-independent, checks
are highly desirable.

Double-diffractive dijet production provides the attractive
possibility to test factorization, in a parameter-free way, using
only data from hadronic collisions. This allows an important
consistency test of the whole approach. Indeed, a test of
factorization has been performed recently by the CDF collaboration
at $\sqrt s = 1.8$~TeV~\cite{CDFjj}. The data of interest are
dijet production in single-diffractive dissociation (SD),
Figs.~1(a),(c), and in the double-Pomeron-exchange (DP) process,
Figs.~1(b),(d). The distributions of the partons\footnote{The Born
diagrams of Figs.~1(a),(b) correspond to the Ingleman--Schlein
conjecture~\cite{IS}.} which collide in Figs.~1(a),(c) to produce
the dijet system may be taken as the effective densities
\be f_a(x)\ \equiv\ g_a(x) + \frac{4}{9}q_a(x)\,,
\label{eq:effectivedensities} \ee
since the hard subprocess is dominated by gluon $t$ channel
exchange. Here $g(x)$ and $q(x)$ denote the gluon and the sum of
quark, antiquark densities, and 4/9 is the appropriate colour
factor. The subscript $a=p$ ($\bar p$) or $P$ indicates whether
the `gluon' belongs to the proton (antiproton) or the Pomeron, as
seen, for example, in the upper and lower parts of Fig.~1(a)
respectively.

Following \cite{CDFjj}, we consider first the ratio of SD dijet
production to non-diffractive (ND) dijet production. Then, in the
ratio of the cross sections (for the same kinematical
characteristics of jets, $E_{Ti}>7~\GeV$), the quantity
$f_p(x_i)\sigma_{jj}(\eta_{j1},\eta_{j2},E_{T1},E_{T2})$ cancels
out, where $\sigma_{jj}$ is the partonic cross section to produce
dijets with pseudorapidities $\eta_{j1}$, $\eta_{j2}$ and
transverse energies $E_{T1}$, $E_{T2}$. Moreover the two processes
are measured by the same detector so, in the ratio, experimental
uncertainties are reduced. Allowing for the above cancellation,
the ratio can be written in the form
\be R_{\rm ND}^{\rm SD}\ \equiv\ \frac{\sigma_{jj}^{\rm
SD}}{\sigma_{jj}^{\rm ND}}\ =\ \frac{F_P(\xi_{\bar
p})f_P(\beta)\,\beta}{f_{\bar p}(x_{\bar p})\,x_{\bar p}}\,S_1\,,
\label{eq:sigmasratio_StoN} \ee
where $F_P(\xi_{\bar p})$ is the Pomeron `flux factor', $\xi_{\bar
p}$ is the fraction of the initial momentum carried by the Pomeron
(experimentally $\xi_{\bar p}<0.1$), $f_P(\beta)$ is the effective
distribution of partons in the Pomeron and $x_{\bar
p}=\beta\xi_{\bar p}$. The suppression factor $S_1$, which,
following Bjorken~\cite{Bj}, is often called the survival
probability, accounts for the screening effects caused by diagrams
of the type shown in Fig.~1(c). It is normalized so that $S_1
\equiv 1$ for the Born diagram of Fig.~1(a).

In order to provide a quantitative check of the
`factorization-violating' suppression factors, it is possible to
remove, to a large extent, the uncertainties associated with the
Pomeron flux factor and the parton distributions and detector
effects, by studying a second ratio of measured cross sections.
Considering the ratio of cross sections for dijet production by
the double-Pomeron-exchange (DP) and the single diffractive (SD)
processes, we have
\be R_{\rm SD}^{\rm DP}\ \equiv\ \frac{\sigma_{jj}^{\rm
DP}}{\sigma_{jj}^{\rm SD}}\ =\
\frac{F_P(\xi_{p})f_P(\beta_1)\,\beta_1}{f_p(x_p)\,
x_p}\,\frac{S_2}{S_1}\,, \label{eq:sigmasratio_DtoS} \ee
where $S_2$ is the suppression factor for the DP process (in
general $S_2\neq S_1$). Then the ratio of the two ratios,
(\ref{eq:sigmasratio_StoN}) and (\ref{eq:sigmasratio_DtoS}),
becomes
\be D\ =\ \frac{R_{\rm ND}^{\rm SD}}{R_{\rm SD}^{\rm DP}}\ =\
\frac{F_P(\xi_{\bar
p})f_P(\beta)\,\beta}{F_P(\xi_{p})f_P(\beta_1)\,\beta_1}\,
\frac{f_p(x_p)\,x_p}{f_{\bar p}(x_{\bar p})\,x_{\bar
p}}\:\frac{S_1^2}{S_2}\,. \label{eq:D_ratioofratios} \ee
In the case when $\xi_{\bar p}=\xi_{p}$ and $\beta = \beta_1$
($x_{\bar p}=x_p$),
 the double ratio becomes
\be D\ =\ \frac{S_1^2}{S_2}\,. \label{eq:D_specialcaseratio} \ee
If there were no multi-Pomeron effects ($S_1=S_2=1$) then $D=1$.
Thus a deviation of $D$ from unity would signal a failure of
factorization. We emphasize that, although we use Regge
factorization (with the cross sections written as products of flux
factors and the corresponding partonic distributions), the result
is practically insensitive to this assumption. The breakdown of
factorization for hard diffractive processes is naturally expected
in QCD; see, for example, \cite{COMPAT,KKMR,C}.

To make a quantitative evaluation of the ratio $D$ we use the
model predictions of Ref.~\cite{KMRsoft} for the multi-Pomeron
screening effects, where the suppression factors $S_i$ were
calculated for a range of hard diffractive processes at the
various $pp$ (and $\pp$) collider energies. For our processes at
$\sqrt s = 1.8$~TeV they were found to be
\be S_1 = 0.10,\quad S_2 = 0.05.
\label{eq:suppressionfactorsat1.8} \ee
The difference between $S_1$ and $S_2$ is due to the difference in
the impact parameter profiles for the processes of diagrams 1(a)
and 1(b). Thus the prediction of the model is $D=0.2$. It was
emphasized in Ref.~\cite{KKMR} that the suppression
factor\footnote{There may also be a $\beta$ dependence of the
suppression factor arising from QCD radiative effects; see, for
example, Ref.~\cite{COMPAT}. However, for the relevant CDF
kinematics with $\beta$ not close to 1, this dependence is not
significant and, moreover, is essentially cancelled out in the
ratio (\ref{eq:D_ratioofratios}).} $S_1$ can depend also on
$x_{\bar p}$ and $\beta$. When more precise data become available,
these dependences can be taken into account. However, in the
kinematical range of the CDF measurement~\cite{CDFjj}, the average
value of $S_1$ turns out to be very close to 0.10.

The experimental value for the double ratio obtained by the CDF
Collaboration~\cite{CDFjj} is $D=0.19\pm0.07$, in good agreement
with the theoretical prediction. It clearly demonstrates the
presence of factorization breaking and the importance of
unitarization effects due to multi-Pomeron exchanges.

It is worth commenting briefly on one aspect of the experimental
determination of the ratio $D$~\cite{CDFjj}. The range of
$\xi_{\bar p}$ covered in the CDF measurements was
$0.035<\xi_{\bar p}<0.095$, while $\xi_{p}$ was in the interval
$0.01<\xi_{p}<0.03$. It was therefore necessary to extrapolate the
second ratio $R_{\rm SD}^{\rm DP}$ in $\xi_{\bar p}$ to the value
$\xi_{\bar p}=0.02$ in order to have the same values of $\xi_i$ in
Eq.~(\ref{eq:D_ratioofratios}). This was done on the assumption
that this ratio is independent of $\xi_{\bar p}$ (for a given
$x_{\bar p}$), as was in fact seen in the data in the observed
range of $\xi_{\bar p}$. This assumption is confirmed by the model
of Ref.~\cite{KKMR}, where a detailed description of the CDF data
on SD dijet production  was performed. It was shown that the
$\xi_{\bar p}$-dependence, arising from the Pomeron and secondary
exchange contributions in the region $0.01\leq\xi_{\bar p}\leq
0.1$, can be approximated by $1/\xi_{\bar p}^{n_1}$ with
$n_1\simeq 1$. Moreover, the $\beta$-dependence in the region
$\beta<0.2$ is also close to $1/\beta^{n_2}$ with $n_2\simeq 1$; a
behaviour which well summarizes several contributing $\beta$
behaviours~\cite{KKMR}. Thus, the total dependence on $\beta$ and
$\xi_{\bar p}$ is $1/\beta\xi_{\bar p} = 1/x_{\bar p}$. Therefore,
for fixed $x_{\bar p}$ there is no dependence on $\xi_{\bar p}$.
The approximate equality of the values of $n_1$ and $n_2$ may not
hold at very small $\xi_{\bar p}$ or at much higher
energies~\cite{KKMR}.

In conclusion, we have demonstrated that a theoretical framework,
which takes into account unitarization effects due to
multi-Pomeron exchanges, predicts a substantial breakdown of
factorization in hard diffractive processes. The multi-Pomeron
effects suppress the Born cross section by {\em different} factors
for {\em different} diffractive processes, due to the {\em
different} impact parameter profiles of the various cross
sections. We have proposed a check of the factorization-violating
suppression factors which depends only on {\em measured} cross
sections at the Tevatron, namely the cross sections for dijet
production in single diffractive and double-Pomeron-exchange
processes. The prediction and the data are in good agreement. A
detailed comparison, along the lines proposed, of forthcoming
precision dijet data from the Tevatron or LHC will be very
informative.

\section*{Acknowledgements}

We thank K.~Goulianos for useful discussions. This work was
partially supported by the UK Particle Physics and Astronomy
Research Council, and by grants INTAS~00-00366,
NATO~PSTCLG-977275, RFBR~00-15-96786, 01-02-17383 and 01-02-17095.

%\newpage

\end{document}